\documentclass[journal,twoside,web]{ieeecolor}
\usepackage{generic}
\usepackage{cite}
\usepackage{amsmath,amssymb,amsfonts}
\usepackage{algorithmic}
\usepackage[pdftex]{graphicx}
\usepackage{textcomp}
\usepackage{siunitx}
\usepackage{subcaption}
\usepackage{hyperref}

\usepackage{mathtools}
\usepackage{float}
\usepackage{booktabs}
\usepackage[bottom]{footmisc}

\usepackage{adjustbox}

\newcommand\ie {{\it i.e.}, }
\newcommand\eg {{\it e.g.}, }

\usepackage{amsmath,amsfonts,bm}

\def\eqref#1{eq.~\ref{#1}}

\def\1{\bm{1}}

\DeclareMathAlphabet{\mathsfit}{\encodingdefault}{\sfdefault}{m}{sl}
\SetMathAlphabet{\mathsfit}{bold}{\encodingdefault}{\sfdefault}{bx}{n}

\newcommand{\E}{\mathbb{E}}

\def\BibTeX{{\rm B\kern-.05em{\sc i\kern-.025em b}\kern-.08em
    T\kern-.1667em\lower.7ex\hbox{E}\kern-.125emX}}
\begin{document}
\title{Hierarchical Amortized Training for Memory-efficient High Resolution 3D GAN}
\author{ Li Sun,  Junxiang Chen,  Yanwu Xu, Mingming Gong,  Ke Yu, and Kayhan Batmanghelich$^{\ast}$
\thanks{Manuscript received Mar 27, 2022.
This work was partially supported by NIH Award Number
1R01HL141813-01, NSF 1839332 Tripod+X, SAP SE and Pennsylvania Department of Health.
We are grateful for the computational
resources provided by Pittsburgh SuperComputing grant number TG-ASC170024. \emph{Asterisk indicates corresponding author.}}
\thanks{ L. Sun,  Y. Xu and  K. Yu are with the 
School of Computing and Information, University of Pittsburgh, Pittsburgh, PA 15206, USA (email: lis118@pitt.edu, yanwuxu@pitt.edu, key44@pitt.edu)}
\thanks{ J. Chen and K. Batmanghelich are with the Department of Biomedical Informatics, University of Pittsburgh, Pittsburgh, PA 15206, USA (email: juc91@pitt.edu, kayhan@pitt.edu)}
\thanks{M. Gong is 
with the School of Mathematics and Statistics, The University of Melbourne, Parkville, VIC, Australia (e-mail: mingming.gong@unimelb.edu.au).}
}

\maketitle

\begin{abstract}
Generative Adversarial Networks (GAN) have many potential medical imaging applications, including data augmentation, domain adaptation, and model explanation. Due to the limited memory of Graphical Processing Units (GPUs), most current 3D GAN models are trained on low-resolution medical images, these models either cannot scale to high-resolution or are prone to patchy artifacts. In this work, we propose a novel end-to-end GAN architecture that can generate high-resolution 3D images.
We achieve this goal by using different configurations between training and inference.
During training, we adopt a hierarchical structure that simultaneously generates a low-resolution version of the image and a randomly selected sub-volume of the high-resolution image. The hierarchical design has two advantages: First, the memory demand for training on high-resolution images is amortized among sub-volumes. Furthermore, anchoring the high-resolution sub-volumes to a single low-resolution image ensures anatomical consistency between sub-volumes. During inference, our model can directly generate full high-resolution images. We also incorporate an encoder with a similar hierarchical structure into the model to extract features from the images. Experiments on 3D thorax CT and brain MRI demonstrate that our approach outperforms state of the art in image generation. We also demonstrate clinical applications of the proposed model in data augmentation and clinical-relevant feature extraction.


\end{abstract}

\begin{IEEEkeywords}
Generative Adversarial Networks, 3D Image Synthesis, High Resolution.
\end{IEEEkeywords}

\section{Introduction}
\label{sec:introduction}

\IEEEPARstart{G}{enerative} Adversarial Networks (GANs) have succeeded in generating realistic-looking natural images ~\cite{goodfellow2014generative,rosca2017variational}. It has shown potential in medical imaging for augmentation~\cite{han1904h2019,shin2018medical}, image reconstruction~\cite{quan2018compressed} and image-to-image translation~\cite{armanious2020medgan,lei2019mri}. 
 The prevalence of 3D images in the radiology domain renders the real-world application of GANs in the medical domain even more challenging than the natural image domain. In this paper, we propose an efficient method for generating and extracting features from high-resolution volumetric images.

The training procedure of GANs corresponds to a min-max game between two players: a generator and a discriminator. While the generator aims to generate realistic-looking images,  the discriminator aims to defeat the generator by recognizing real from the fake (generated) images.
When the field of view (FOV) is the same, a higher resolution is equivalent to more voxels. In this way, we use ``high-resolution image" and ``large-size image" interchangeably in the paper.
In clinical application, radiologists rely on high-resolution CT to make accurate diagnose decisions~\cite{bonelli1998accuracy}.
While there are previous works that propose to use 3D GAN for diverse medical applications~\cite{cirillo2020vox2vox,yu20183d}, the generated images are limited to the small size of $128 \times 128 \times 128$ or below, due to insufficient memory during training.

In this paper, we introduce a Hierarchical Amortized GAN (HA-GAN) to bridge the gap. 
Our model adopts different configurations between training and inference phases.
In the training phase, we simultaneously generate a low-resolution image and a randomly selected sub-volume of the high-resolution image. Generating sub-volumes amortizes the memory cost of the high-resolution image and keeps local details of the 3D image. Furthermore, the low-resolution image ensures anatomical consistency and the global structure of the generated images. We train the model in an end-to-end fashion while retaining memory efficiency. 
The gradients of the parameters, which are the memory bottleneck, are needed only during training. Hence, sub-volume selection is no longer needed and the entire high-resolution volume can be generated during inference.
In addition, we implement an encoder in a similar fashion. The encoder enables us to extract features from a given image and prevents the model from mode collapse. We test HA-GAN on thorax CT and brain MRI datasets. Experiments demonstrate that our approach outperforms baselines in image generation. We also present two clinical applications with proposed HA-GAN, including data augmentation for supervised learning and clinical-relevant feature extraction. Our code is publicly available at \href{https://github.com/batmanlab/HA-GAN}{https://github.com/batmanlab/HA-GAN} 

In summary, we make the following contributions:

\begin{enumerate}
\item We introduce a novel end-to-end HA-GAN architecture that can generate high-resolution volumetric images while being memory efficient.
\item We incorporate a memory-efficient encoder with a similar structure, enabling clinical-relevant feature extraction from high-resolution 3D images. We show that the encoder improves generation quality.
 \item We discover that moving along specific directions in latent space results in explainable anatomical variations in generated images.
\item We evaluate our method by extensive experiments on different image modalities as well as different anatomy. The HA-GAN offers significant quantitative and qualitative improvements over the state-of-the-art.
\end{enumerate}

\section{Related work}
\label{sec:related_work}
In the following, we review the works related to GANs for medical images, memory-efficient 3D GAN and representation learning in generative models.

\subsection{ GANs for Medical Imaging }
In recent years, researchers have developed GAN-based models for medical images. These models are applied to solve various problems, including image synthesis~\cite{chuquicusma2018fool}, data augmentation~\cite{frid2018synthetic},
modality/style transformation~\cite{zhao2017synthesizing}, and model explanation ~\cite{singla2019explanation}. However, most of these methods concentrate on generating 2D medical images. In this paper, we focus on solving a more challenging problem, i.e., generating 3D images.

With the prevalence of 3D imaging in medical applications, 3D GAN models have become a popular research topic. Shan et al. \cite{shan20183} proposed a 3D conditional GAN model for low-dose CT denoising.  Kudo et al. \cite{simo2019virtual} proposed a 3D GAN model for CT image super-resolution. Jin et al. \cite{jin2019applying} propose an auto-encoding GAN for generating 3D brain MRI images. Cirillo et al. \cite{cirillo2020vox2vox} proposed to use a 3D model conditioned on multi-channel 3D Brain MR images to generate tumor masks for segmentation. While these methods can generate realistic-looking 3D MRI or CT images, the generated images are limited to the small size of $128\times128\times128$ or below, due to insufficient memory during training. 
In contrast, our HA-GAN is a memory-efficient model and can generate 3D images with a size of $256\times256\times256$.

\subsection{Memory-Efficient GANs}
Some works are proposed to reduce the memory demand of high-resolution 3D image generation. In order to address the memory challenge, some works adopt slice-wise~\cite{lei2019mri} or patch-wise~\cite{yu20183d} generation approach. Unfortunately, these methods may introduce artifacts at the intersection between patches/slices because they are generated independently. To remedy this problem, Uzunova et al.~\cite{uzunova2019multi} propose a multi-scale approach that uses a GAN model to generate a low-resolution version of the image first. An additional GAN model is used to generate higher resolution patches of images conditioned on the previously generated patches of lower resolution images. However, this method is still patch-based; the generation of local patches is unaware of the global structure, potentially leading to spatial inconsistency. In addition, the model is not trained in an end-to-end manner, which makes it challenging to incorporate an encoder that learns the latent representations for the entire images. In comparison, our proposed HA-GAN is global structure-aware and can be trained end-to-end. This allows HA-GAN to be associated with an encoder.

\subsection{Representation Learning in Generative Models}
Several existing generative models are fused with an encoder~\cite{diederik2014auto,rosca2017variational,donahue2016adversarial}, which learns meaningful representations for images. These methods are based on the belief that a good generative model that reconstructs realistic data will automatically learn a meaningful representation of it~\cite{chen2016infogan}. A generative model with an encoder can be regarded as a compression algorithm \cite{Townsend2020HiLLoC}. Hence, the model is less likely to suffer from mode collapse because the decoder is required to reconstruct all samples in the dataset, which is impossible if mode collapse happens such that only limited varieties of samples are generated\cite{rosca2017variational}. 
Variational autoencoder (VAE) \cite{diederik2014auto} uses an encoder to compress data into a latent space, and a decoder is used to reconstruct the data using the encoded representation. BiGAN \cite{donahue2016adversarial} learns a bidirectional mapping between data space and latent space.
$\alpha$-GAN \cite{rosca2017variational} introduces not only an encoder to the GAN model, but also learns a disentangled representation by implementing a code discriminator, which forces the distribution of the code to be indistinguishable from that of random noise. Variational auto-encoder GAN (VAE-GAN) \cite{larsen2016autoencoding} adds an adversarial loss to the variational evidence lower bound objective.
Despite their success, the methods mentioned above can analyze 2D images or low-resolution 3D images, which are less memory intensive for training an encoder. In contrast, our proposed HA-GAN is memory efficient and can be used to encode and generate high-resolution 3D images during inference.

\subsection{Our previous work}
Sun \emph{et al.}~\cite{sun2020workshop} first 
proposed to
utilize hierarchical amortized GAN for high resolution 3D medical image generation.
The current work presents several
extensions compared to the preliminary version:
1) We incorporate a memory-efficient encoder into our model, enabling clinical-relevant feature extraction from high-resolution 3D images. We also show that the encoder improves generation quality.
2) We perform two new clinical applications, including characterizing the severity of COPD, and data augmentation for supervised learning.
3) We discover that moving along specific directions in latent space results in explainable anatomical variations in generated images.
4) We perform cross-validation evaluation and statistical tests for comparison of generated image quality with baseline methods to improve the adequacy of performance evaluation. We also conduct ablation studies to validate the contribution of proposed components.

\section{Method}

We first review Generative Adversarial Networks (GANs) in Section~\ref{sec:background}. Then, we introduce our method in Section~\ref{sec:hierarchical}, followed by the introduction of the encoder in Section~\ref{sec:encoder}. We conclude this section with the optimization scheme in Section~\ref{sec:overall} and the implementation details in Section~\ref{sec:implementation}. The notations used are summarized in Table \ref{tbl:notation}.

 \begin{table}[htp]
 {
 \caption{Important notations in this paper}
 \scriptsize
 \begin{flushleft}
 \begin{minipage}[]{.25\textwidth}
 \begin{tabular}{r   l}
     \toprule
     \multicolumn{2}{l}{
     \textbf{Models}} \\
     \midrule
     $G^A(\cdot)$ & The common block of the generator.  \\
     $G^L(\cdot)$ & The low-resolution block of the generator. \\ 
     $G^H(\cdot)$ & The high-resolution block of the generator.\\
     $D^H(\cdot)$ & The discriminator for high-resolution images.\\
     $D^L(\cdot)$ & The discriminator for low-resolution images. \\
     $E^H(\cdot)$ & The high-resolution block of the encoder. \\
     $E^G(\cdot)$ & The ground block of the encoder. \\
     \midrule
     \multicolumn{2}{l}{\textbf{Functions}} \\
     \midrule
     $S^H(\cdot, \cdot)$ & The high-resolution sub-volume selector.\\
     $S^L(\cdot, \cdot)$ & The low-resolution sub-volume selector.\\
     \midrule
     \multicolumn{2}{l}{\textbf{Variables}} \\
     \midrule
     $Z$ & Latent representations.\\
     $\widehat{Z}$ & Reconstructed latent representations.\\
     $c$ & GOLD score. \\
     $r$ & The index of the tarting slice  for sub-volume selection.\\
     $X^H$ & The real high-resolution image.\\
     $X^L$ & The real low-resolution image.\\
     $\widehat{X}^H$ & The generated high-resolution image.\\
     $\widehat{X}^H_r$ & The generated high-resolution sub-volume starting at slice $r$.\\
     $\widehat{X}^L$ & The generated low-resolution image.\\
     $A$ & Intermediate feature maps for the whole image\\
     $A_r$ & Intermediate feature maps for the sub-volume starting at slice $r$ \\
     $\widehat{A}$ & Reconstructed intermediate feature maps for the whole image.\\
     $\widehat{A}_v$ & Reconstructed intermediate feature maps for the $v$-th sub-volume.\\
     $\{T_v\}_{v=1}^V$ & The indices of the starting slices for a partition for $X^H$.
     \\     
    \bottomrule
  \end{tabular} 
 \end{minipage}
 \end{flushleft}

 \label{tbl:notation}
 }
 \vspace{-10mm}
 \end{table}

\subsection{Background}
\label{sec:background}
Generative Adversarial Networks (GANs)\cite{goodfellow2014generative} is widely used to generate realistic-looking images. The training procedure of GANs corresponds to a two-player game that involves a generator $G$ and a discriminator $D$. In the game, while $G$ aims to generate realistic-looking images, $D$ tries to discriminate real images from the images synthesized by $G$. The $D$ and $G$ compete with each other.
Let $P_X$ denote the underlying data distribution, and $P_Z$ denote the distribution of the random noise $Z$. Then the objective of GAN is formulated as below: 
\begin{IEEEeqnarray}{c}
\min_G\max_D\underset{X \sim P_{X}}{\E} [\log D(X)]+\underset{Z\sim P_Z}{\E} [\log(1- D(G(Z)))].
\label{Eq:gan}
\end{IEEEeqnarray}


\begin{figure*}[t]
    \centering
    \includegraphics[width = \textwidth]{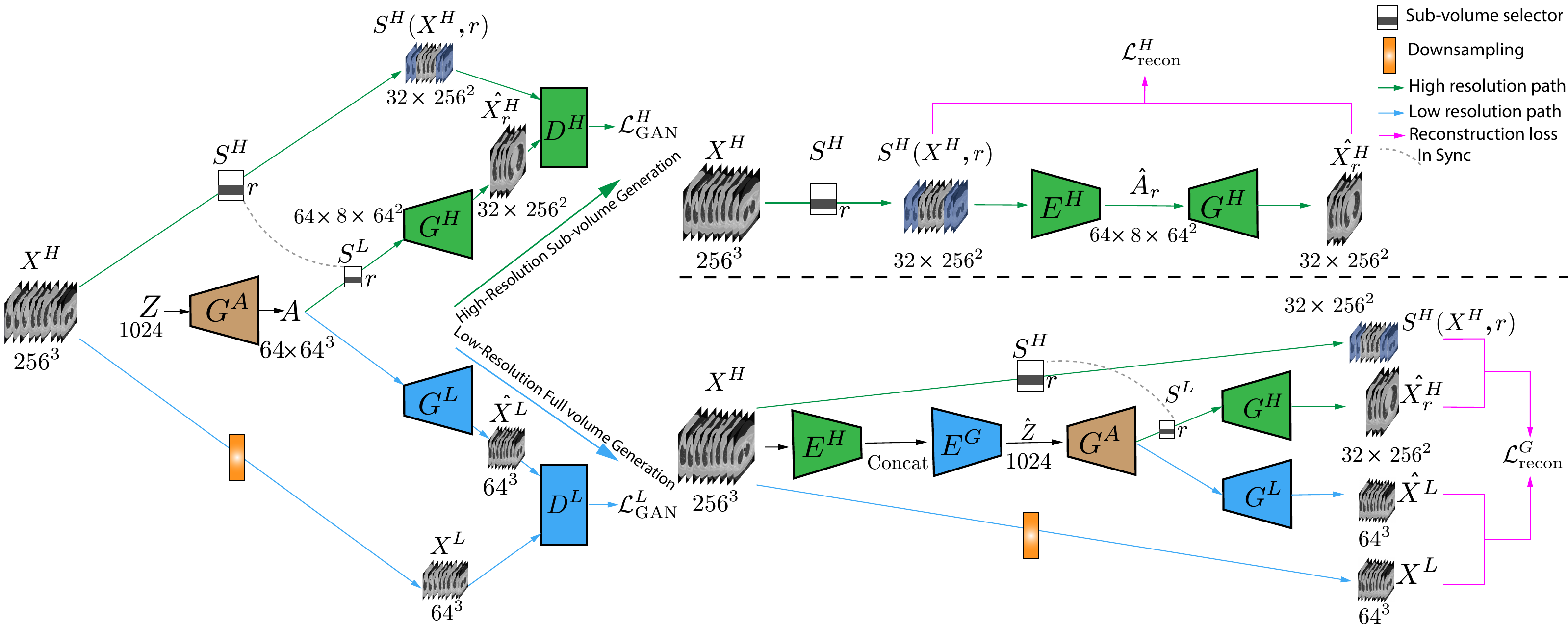}
    \caption{ \textbf{Left:} The architecture of HA-GAN (encoder is hidden here to improve clarity). At the training time, instead of directly generating high-resolution full volume, our generator contains two branches for high-resolution sub-volume and low-resolution full volume generation, respectively. The two branches share the common block $G^A$. A sub-volume selector is used to select a part of the intermediate feature for the sub-volume generation.
    \textbf{Right:} The schematic of the hierarchical encoder trained with two reconstruction losses, one on the high-resolution sub-volume level (upper right) and another one on the low-resolution full volume level (lower right). The meanings of the notations used can be found in Table~\ref{tbl:notation}. The model adopts 3D architecture with details presented in Supplementary Material.}
    \label{fig:GAN_loss}
    \vspace{-3mm}
\end{figure*}

\begin{figure}[t]
    \centering
    \includegraphics[width = .45\textwidth]{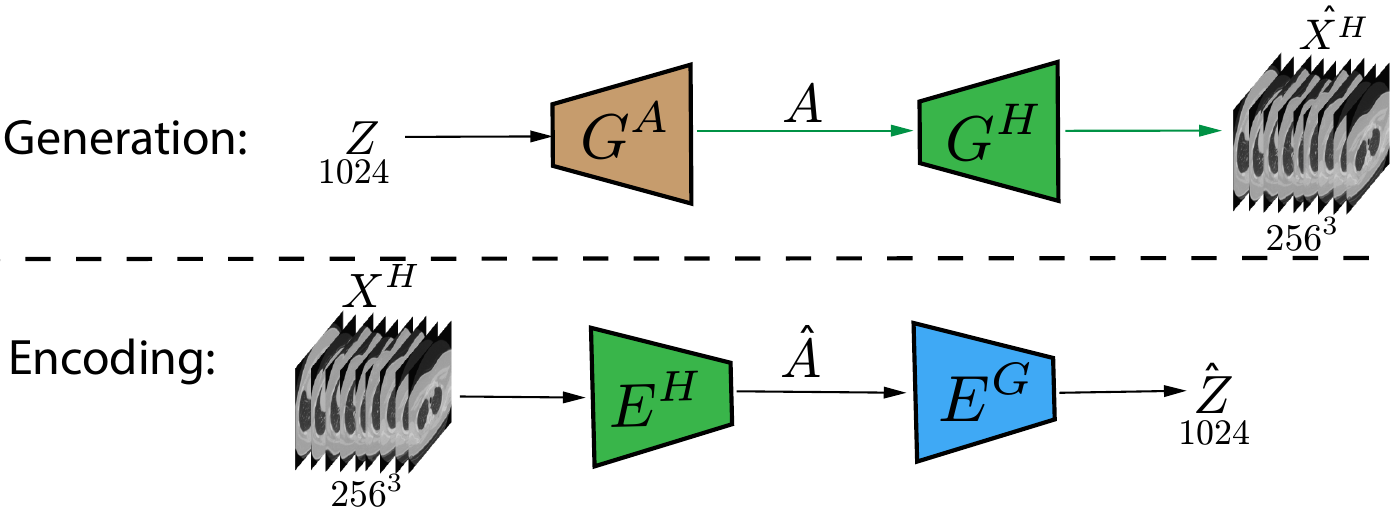}
    \caption{ Inference with the hierarchical generator and encoder. Since the memory demand is lower at inference time, we directly forward input through the high-resolution branch for full image generation and encoding.}
    \label{fig:inference}
    \vspace{-3mm}
\end{figure}

\subsection{The Hierarchical Structure}
\label{sec:hierarchical}
{\vspace{2mm} \noindent \bf Generator \hspace{2mm}}  Our generator has two branches that generate the low-resolution image $\widehat{X}^L$ and a randomly selected sub-volume of the high-resolution image $\widehat{X}^H_r$, where $r$ represents the index  for  the  starting  slice of the sub-volume. The two branches share initial layers $G^A$ and after they branch off:
\begin{IEEEeqnarray}{ll}
    \widehat{X}^L &= G^L (\  \underbrace{ G^A(Z) }_{A} \  ),\\
    \widehat{X}^H_r &= G^H( \  \underbrace{ S^L(G^A(Z);r) }_{A_r} \ ),
    \label{equ:A_r}
\end{IEEEeqnarray}
where $G^A(\cdot)$, $G^L(\cdot)$ and $G^H(\cdot)$ denote the common, low-resolution and high-resolution blocks of the generator, respectively. $S^L(\cdot, r)$ is a selector function that returns the sub-volume of input image starting at slice $r$, where the superscript $L$ indicates that the selection is made at low resolution. The output of this function is fed into $G^H(\cdot)$, which lifts the input to the high resolution. We use $A$ and $A_r$ as short-hand notation for $G^A(Z)$ and $S^L(G^A(Z);r)$, respectively. We let $Z \sim \mathcal{N}( \mathbf{0}, \mathbf{I})$ be the input random noise vector. We let $r$ be the randomly selected index for the starting slice that is drawn from a uniform distribution, denoted as $r \sim \mathcal{U}$; i.e., each slice is selected with the same probability.  
Therefore, the randomly selected sub-volumes can be overlapping, which can better cover the junctions between sub-volumes than non-overlapping sub-volume selection.
  The schematic of the proposed method is shown in Fig.~\ref{fig:GAN_loss}. Note that $\widehat{X}^H_r$ depends on a corresponding sub-volume of $A$, which is $A_r$. Therefore, we feed $A_r$ rather than complete $A$ into $G^H$ during training, making the model memory-efficient.


{\vspace{2mm} \noindent \bf Discriminator \hspace{2mm}}  Similarly, we define two discriminators $D^H$ and $D^L$ to distinguish a real high-resolution sub-volume $X^H_r$ and a low-resolution image $X^L$ from the fake ones, respectively. $D^H$ makes sure that the local details in the high-resolution sub-volume look realistic. At the same time, $D^L$ ensures the proper global structure is preserved. Since we feed a sub-volumes $S^H(X^H;r)$ rather than the entire image $X^H$ into $D^H$, the memory cost of the model is reduced. The location of the sub-volume $r$ is also fed into $D^H$ to help it distinguish sub-volumes from different locations.  

There are two GAN losses $\mathcal{L}^H_{GAN}$ and $\mathcal{L}^L_{GAN}$ for high and low resolutions respectively:

\begin{equation}
\footnotesize
\begin{aligned}
\mathcal{L}^H_{GAN}(G^A,G^H,D^H)
= & \underset{G^H,G^A}{\min} \underset{D^H}{\max} \underset{r\sim U}{\E}
\left[ \underset{X \sim P_{X}}{\E} [\log D^H( S^H(X^H;r), r )]
\right.
\\
& +
\left.
\underset{Z\sim P_Z}{\E} [\log(1- D^H( \widehat{X}^H_r,r )]
\right] , 
\end{aligned}
\vspace{-5mm}
\label{equ:L_H}
\end{equation}

\begin{equation}
\footnotesize
\begin{aligned}
\mathcal{L}^L_{GAN}(G^L,G^A,D^L) 
= &
\underset{G^L,G^A}{\min} \underset{D^L}{\max}
\underset{X \sim P_{X}}{\E}[\log D^L(X^L)]
\\
& +\underset{z\sim P_Z}{\E} [\log(1- D^L(\widehat{X}^L)].
\end{aligned}
\end{equation}
Note that the sampler $S^H(\cdot;r)$ in Equation (\ref{equ:A_r}) and $S^L(\cdot;r)$ in Equation (\ref{equ:L_H}) are synchronized, such that $r$  corresponds to the indices for the same percentile of slices in the high- and low-resolution.

{\vspace{2mm} \noindent \bf Inference \hspace{2mm}}  
The memory space needed to store gradient is the main bottleneck for 3D GANs models; however, the gradient is not needed during inference. Therefore, we can directly generate the high-resolution image by feeding $Z$ into $G^A$ and $G^H$ sequentially, \ie $\widehat{X}^H(Z) = G^H( G^A(Z) )$). Note that to generate the entire image during inference, we directly feed the complete feature maps $A = G^A(Z)$ rather than its sub-volume $A_r$ into the convolutional network $G^H$.

\subsection{Incorporating the Encoder}
\label{sec:encoder}
We also adopt a hierarchical structure for the encoder, by defining two encoders $E^H(\cdot)$ and $E^G(\cdot)$ encoding the high-resolution sub-volume and the entire image respectively. We partition the high-resolution image $X^H$ into a set of $V$ \emph{non-overlapping} sub-volumes, 
\ie $X^H = \texttt{concat}(\{ S^H(X^H, T_v) \}_{v=1}^V$), where $\texttt{concat}$ represent concatenation, $S^H(\cdot)$ represents the selector function that returns a sub-volume of a high-resolution image, and $T_v$ represents the corresponding starting indices for the non-overlapping partition. 

We use $\widehat{A}_v$ to denote the sub-volume-level feature maps for the $v$-th sub-volume, i.e., $\widehat{A}_v = E^H( S^H(X^H; T_v))$. To generate the image-level representation $\widehat{Z}$, we first summarize all sub-volume representation for the image through concatenation, such that $\widehat{A} = \texttt{concat}(\{A_{v}\}_{v=1}^V)$. Then we feed $\widehat{A}$ into the encoder $E^G(\cdot)$ to generate the image-level representation $\widehat{Z}$, i.e., 
$\widehat{Z} = E^G(\widehat{A})$


 In order to obtain optimal $E^H$ and $E^G$, we introduce the following objective functions:
\begin{equation}
    \label{Eq:recon_H}
    \small
    \mathcal{L}^H_{recon}(E^H)=
    \min_{E^H}
    \underset{X \sim P_{X}, r \in U } {\E} \left\lVert S^H(X^H;r) - G^H( \widehat{A}_r )\right\rVert_{1},
\end{equation}
\begin{equation}
\label{Eq:recon_G}
\resizebox{.9\linewidth}{!}{$
\begin{aligned}
\mathcal{L}^G_{recon}(E^G) = & \underset{E^G}{\min}
\underset{X \sim P_{X}}{\E} 
\left[ 
\left\lVert X^L - G^L(G^A(\widehat{Z}))\right\rVert_{1}
\right.
\\
& + \left.
\underset{r \sim U}{\E} 
\left[ 
\left\lVert S^H(X^H;r)
- G^H(S^L(G^A(\widehat{Z});r) )\right\rVert_{1}
\right] 
\right].
\end{aligned}
$}
\end{equation}
Equation~(\ref{Eq:recon_H}) ensures a randomly selected high-resolution sub-volume $S^H(X^H;r)$ can be reconstructed. Equation~(\ref{Eq:recon_G}) enforces both the low-resolution image $X^L$ and a random selected $S^H(X^H;r)$ can be reconstructed given $\widehat{Z}$. Note that in Equation~(\ref{Eq:recon_H}), the sub-volume is reconstructed from the intermediate feature maps $\widehat{A}_v$; while in the second term in Equation (\ref{Eq:recon_G}), the sub-volume is reconstructed from the latent representations $\widehat{Z}$. In these equations, we use $\ell_1$ loss for reconstruction because it tends to generate sharper result compared to $\ell_2$ loss~\cite{zhu2017unpaired}. The structure of the encoders are illustrated in Fig.~\ref{fig:GAN_loss}. 

When optimizing for Equation~(\ref{Eq:recon_H}), we only update $E^H$ while keeping all other parameters fixed. Similarly, when optimizing for Equation~(\ref{Eq:recon_G}), we only update $E^G$. We empirically find that this optimization strategy is memory-efficient and leads to better performance.

{\vspace{2mm} \noindent \bf Inference \hspace{2mm}} In the inference phase, we can get the latent code $\widehat{Z}$ by feeding the sub-volumes of $X^H$ into $E^H$, concatenating the output sub-volume feature maps into $\widehat{A}$ and then feeding the results into $E^G$, \ie 
$\widehat{Z} = E^G ( \texttt{concat} ( \{ E^H( S^H(X^H; T_v) ) \}_{v=1}^V ) ) $. The idea is illustrated at the bottom of Fig.~\ref{fig:inference}.

\subsection{Overall Model}
\label{sec:overall}
The model is trained in an end-to-end fashion. The overall loss function is defined as:
\begin{equation}
\begin{aligned}
    \mathcal{L} & = 
    \mathcal{L}^H_{GAN}(G^H,G^A,D^H) + 
    \mathcal{L}^L_{GAN}(G^L,G^A,D^L)
    \\
    & + \lambda_1 \mathcal{L}^H_{recon}(E^H) +
    \lambda_2 \mathcal{L}^G_{recon}(E^G),
\end{aligned}
\end{equation}
where $\lambda_1$ and $\lambda_2$ control the trade-off between the GANs losses and the reconstruction losses. The optimizations for generator ($G^H$, $G^L$ and $G^A$), discriminator ($D^H$, $D^L$), and encoder ($E^H$, $E^G$) are altered per iteration.

During training, we sample noise from Gaussian distribution and pass it through the generator to create randomly synthesized images for minimizing the adversarial loss. We also sample real images and pass it through the encoder, followed by the generator to create reconstructed images for minimizing the reconstruction loss. Our overall optimization balances between the losses to learn parameters for the encoder, generator, and discriminator in end-to-end training.

\subsection{Implementation Details}
\label{sec:implementation}
We train the proposed HA-GAN for 80000 iterations, the training and validation curves can be found in Supplementary Material. We let the learning rate for generator, encoder, and discriminator be $1\times10^{-4}$, $1\times10^{-4}$, and $4\times10^{-4}$, respectively. We also set $\beta_1 = 0$ and $\beta_2 = 0.999$ for the Adam optimizer. The batch size is set as 4. We let the size of the $X^L$ be $64^3$. The size of the randomly selected sub-volume $S^H(X^H;r)$ is defined to be $32 \times 256^2$, where $r$ is randomly selected on the batch level. We let feature maps $A$ have $64$ channels with a size of $64^3$. The dimension of the latent variable $Z$ is chosen to be 1,024. The trade-off hyper-parameters $\lambda_1$ and $\lambda_2$ are set to be 5. The experiments are performed on two NVIDIA Titan Xp GPUs, each with 12GB GPU memory. The detailed architecture can be found in Supplementary Material.


\section{Experiments}
We evaluate the proposed model's performance in  image synthesis, and demonstrate two clinical applications with HA-GAN: data augmentation and clinical-relevant feature extraction. We also explore the semantic meaning of the latent variable. We perform 5-fold cross-validation for the image synthesis experiments. We compare our method with baseline methods, including WGAN~\cite{gulrajani2017improved}, VAE-GAN~\cite{larsen2016autoencoding}, $\alpha$-GAN~\cite{kwon2019generation}, Progressive GAN~\cite{progressive_gan}, 3D StyleGAN 2~\cite{hong20213d} and CCE-GAN~\cite{xing2021cycle}.

\subsection{Datasets}

The experiments are conducted on two large-scale 3D datasets, including the COPDGene dataset~\cite{regan2011genetic} and the GSP dataset~\cite{holmes2015brain}. Both are publicly available and details about image acquisition are presented in Supplementary Material.

{\vspace{2mm} \par \bf \noindent COPDGene Dataset: \hspace{2mm}}
We use 3D thorax computerized tomography (CT) images of 9,276 subjects from COPDGene dataset in our study. Only full inspiration scans are used in our study. We trim blank axial slices with all-zero values and resize the images to $256^3$. The Hounsfield units of the CT images have been calibrated and air density correction has been applied. The Hounsfield Units (HU) are mapped to the intensity window of $[-1024,600]$ and normalized to $[-1,1]$.

{\vspace{2mm} \par \bf \noindent GSP Dataset: \hspace{2mm} }
We use 3D Brain magnetic resonance images (MRIs) of 3,538 subjects from the Brain Genomics Superstruct Project (GSP)~\cite{holmes2015brain} in our experiments. The FreeSurfer package~\cite{fischl2012freesurfer} is used to remove the non-brain region in the images, bias-field correction, intensity normalization, affine registration to Talairach space, and resampling to $1 mm^3$
isotropic resolution. We trim the blank axial slices with all-zero values and rescale the images into $256^3$. The intensity value is clipped at top $0.1\%$ quantile to remove outliers, and then normalized into $[-1,1]$.

\subsection{Image Synthesis}
We examine whether the synthetic images are realistic-looking quantitatively and qualitatively, where synthetic images are generated by feeding random noise into the generator.

\subsubsection{Quantitative Evaluation}
If the synthetic images are realistic-looking, then the synthetic images' distribution should be indistinguishable from that of the real images. Therefore, we can quantitatively evaluate the quality of the synthetic images by Fréchet Inception Distance (FID)~\cite{heusel2017gans}, Maximum Mean Discrepancy (MMD)~\cite{gretton2012kernel} and Inception Score (IS)~\cite{salimans2016improved}. Lower values of FID/MMD and higher values of IS indicate that the distributions of generated images are closer to real ones, implying more realistic-looking synthetic images. We evaluate the synthesis quality at two resolutions: $128^3$ and $256^3$. Due to memory limitations, the baseline models can only be trained with the size of $128^3$ at most. To make a fair comparison with our model (HA-GAN), we apply trilinear interpolation to upsample the synthetic images of baseline models to $256^3$.
We adopt a 3D ResNet model pre-trained on 3D medical images~\cite{chen2019med3d} to extract features for computing FID and MMD. Note the scale of FID relies on the feature extraction model. Thus our FID values are not comparable to FID value calculated on 2D images, which is based on feature extracted using model pre-trained on ImageNet. 
For the IS scores, following the practice of~\cite{hong20213d}, we measure the Inception Scores on the middle slices on axial,
coronal, and sagittal planes of the generated 3D images and report averaged performance.
As shown in Table~\ref{tbl:FID_COPD} and Table~\ref{tbl:FID_GSP}, HA-GAN achieves lower FID and MMD as well as higher IS than the baselines, which implies that HA-GAN generates more realistic images. We found that at the resolution of $128^3$, HA-GAN still outperforms the baseline models, but the lead has been smaller compared with the result at the resolution of $256^3$. 
In addition, we performed statistical tests on the evaluation results at $256^3$ resolution between methods. More specifically, we performed two-sample $t$-tests (one-tailed) between HA-GAN and each of the baseline methods. At a significance level of 0.05, HA-GAN achieves significantly higher performance than baseline methods for both datasets.

\begin{table*}[htp]
 \centering
 \caption{ Evaluation for image synthesis on COPDGene dataset}
 \begin{adjustbox}{max width=\textwidth}
 \label{tbl:FID_COPD}
  \begin{tabular}{lccc|ccc}
  \toprule
   Resolution&\multicolumn{3}{c|}{$128^3$}&\multicolumn{3}{c}{$256^3$} \\
   \toprule
     &  FID$\downarrow$&  MMD$\downarrow$& IS$\uparrow$& FID$\downarrow$&  MMD$\downarrow$ & IS$\uparrow$ \\
   \midrule
 WGAN&       $0.012_{\pm.011}$&$0.092_{\pm.059}$&$1.99_{\pm.07}$&
             $0.161_{\pm.044}$&$0.471_{\pm.110}$&$1.97_{\pm.05}$
              \\
 VAE-GAN&    $0.139_{\pm.002}$&$1.065_{\pm.008}$&$1.19_{\pm.03}$&
             $0.328_{\pm.007}$&$1.028_{\pm.008}$&$1.18_{\pm.03}$
              \\
 $\alpha$-GAN& $0.010_{\pm.004}$&$0.089_{\pm.056}$&$1.89_{\pm.04}$&
               $0.043_{\pm.094}$&$0.323_{\pm.080}$&$1.96_{\pm.03}$
               	 \\
 Progressive GAN&$0.015_{\pm.007}$&$0.150_{\pm.072}$&$1.75_{\pm.11}$&
                 $0.107_{\pm.037}$&$0.287_{\pm.123}$&$1.76_{\pm.11}$
                 \\
StyleGAN 2&$0.011_{\pm.001}$&$0.071_{\pm.002}$&$2.03_{\pm.02}$&
           $0.081_{\pm.003}$&$0.225_{\pm.008}$&$2.06_{\pm.01}$
                 \\
CCE-GAN &$0.010_{\pm.004}$&$0.087_{\pm.039}$&$1.97_{\pm.05}$&
         $0.074_{\pm.038}$&$0.252_{\pm.116}$&$1.95_{\pm.04}$
                 \\
 \midrule
 HA-GAN & \bm{$0.005_{\pm.003}$}&\bm{$0.038_{\pm.020}$}&\bm{$2.05_{\pm.05}$}&
          \bm{$0.008_{\pm.003}$}&\bm{$0.022_{\pm.010}$}&\bm{$2.09_{\pm.06}$}
                \\
  \bottomrule
  \end{tabular}
  \vspace{-3mm}
  \end{adjustbox}
 \end{table*}
 
 \begin{table*}[htp]
 \centering
 \caption{ Evaluation for image synthesis on GSP dataset}
 \begin{adjustbox}{max width=\textwidth}
 \label{tbl:FID_GSP}
  \begin{tabular}{lccc|ccc}
  \toprule
   Resolution&\multicolumn{3}{c|}{$128^3$}&\multicolumn{3}{c}{$256^3$} \\
   \toprule
     &  FID$\downarrow$&  MMD$\downarrow$& IS$\uparrow$& FID$\downarrow$&  MMD$\downarrow$ & IS$\uparrow$ \\
   \midrule
 WGAN&       $0.006_{\pm.002}$&$0.406_{\pm.143}$&$1.37_{\pm.02}$&
             $0.025_{\pm.013}$&$0.328_{\pm.139}$&$1.43_{\pm.03}$
              \\
 VAE-GAN&    $0.075_{\pm.004}$&$0.667_{\pm.026}$&$1.03_{\pm.01}$&
             $0.635_{\pm.040}$&$0.702_{\pm.028}$&$1.06_{\pm.06}$
              \\
 $\alpha$-GAN& $0.010_{\pm.007}$&$0.606_{\pm.204}$&$1.39_{\pm.03}$&
               $0.029_{\pm.016}$&$0.428_{\pm.141}$&$1.34_{\pm.08}$	 \\
 Progressive GAN&$0.017_{\pm.008}$&$0.818_{\pm.217}$&$1.25_{\pm.10}$&
                 $0.127_{\pm.055}$&$1.041_{\pm.239}$&$1.25_{\pm.10}$\\
StyleGAN 2&$0.014_{\pm.001}$&$0.369_{\pm.175}$&$1.26_{\pm.01}$&
           $0.048_{\pm.001}$&$0.370_{\pm.020}$&$1.32_{\pm.01}$\\
CCE-GAN &$0.005_{\pm.004}$&$0.301_{\pm.147}$&$1.38_{\pm.02}$&
         $0.030_{\pm.011}$&$0.411_{\pm.106}$&$1.41_{\pm.04}$\\
 \midrule
 HA-GAN &  \bm{$0.002_{\pm.001}$}&\bm{$0.129_{\pm.026}$}&\bm{$1.41_{\pm.02}$}&
           \bm{$0.004_{\pm.001}$}&\bm{$0.086_{\pm.029}$}&\bm{$1.50_{\pm.03}$} \\
  \bottomrule
  \end{tabular}
  \vspace{-2mm}
  \end{adjustbox}
 \end{table*}

\subsubsection{Ablation Study}
We perform three ablation studies to validate the contribution of each of the proposed components. The experiments are performed at $256^3$ resolution. Shown in Table~\ref{tbl:ablation}, we found that adding a low-resolution branch can help improve results, since it can help the model learn the global structure. Adding an encoder can also help improve performance, since it can help stabilize the training. For the deterministic $r$ experiments, we make the sub-volume selector to use a set of deterministic values of $r$ (equal interval between them) rather than the randomly sampled $r$ currently used. From the results, we can see that randomly sampled $r$ outperforms deterministic $r$.

\begin{table*}[htp]
 \centering
 \vspace{-1mm}
 \caption{ Results of ablation study}
 \begin{adjustbox}{max width=\textwidth}
 \label{tbl:ablation}
  \begin{tabular}{lcccc}
  \toprule
   Dataset&\multicolumn{2}{c}{COPDGene (Lung)} & \multicolumn{2}{c}{GSP (Brain)}\\
   \toprule
     &  FID$\downarrow$&  MMD$\downarrow$&  FID$\downarrow$&  MMD$\downarrow$ \\
   \midrule
 HA-GAN w/o Low-resolution branch& $0.030_{\pm.018}$&$0.071_{\pm.039}$&
                            $0.118_{\pm.078}$&$0.876_{\pm.182}$ \\
 HA-GAN w/o Encoder& $0.010_{\pm.003}$&$0.034_{\pm.006}$&
                            $0.006_{\pm.003}$&$0.099_{\pm.028}$ \\
  HA-GAN w/ Deterministic $r$ & $0.014_{\pm.003}$&$0.035_{\pm.007}$&
                            $0.061_{\pm.016}$&$0.612_{\pm.157}$ \\   
 HA-GAN &      \bm{$0.008_{\pm.003}$}&\bm{$0.022_{\pm.010}$}&
               \bm{$0.004_{\pm.001}$}&\bm{$0.086_{\pm.029}$} \\
  \bottomrule
  \end{tabular}
  \vspace{-6mm}
  \end{adjustbox}
 \end{table*}

\subsubsection{Qualitative Evaluation}
\begin{figure*}[t!]
    \centering
    \includegraphics[width = \textwidth]
    {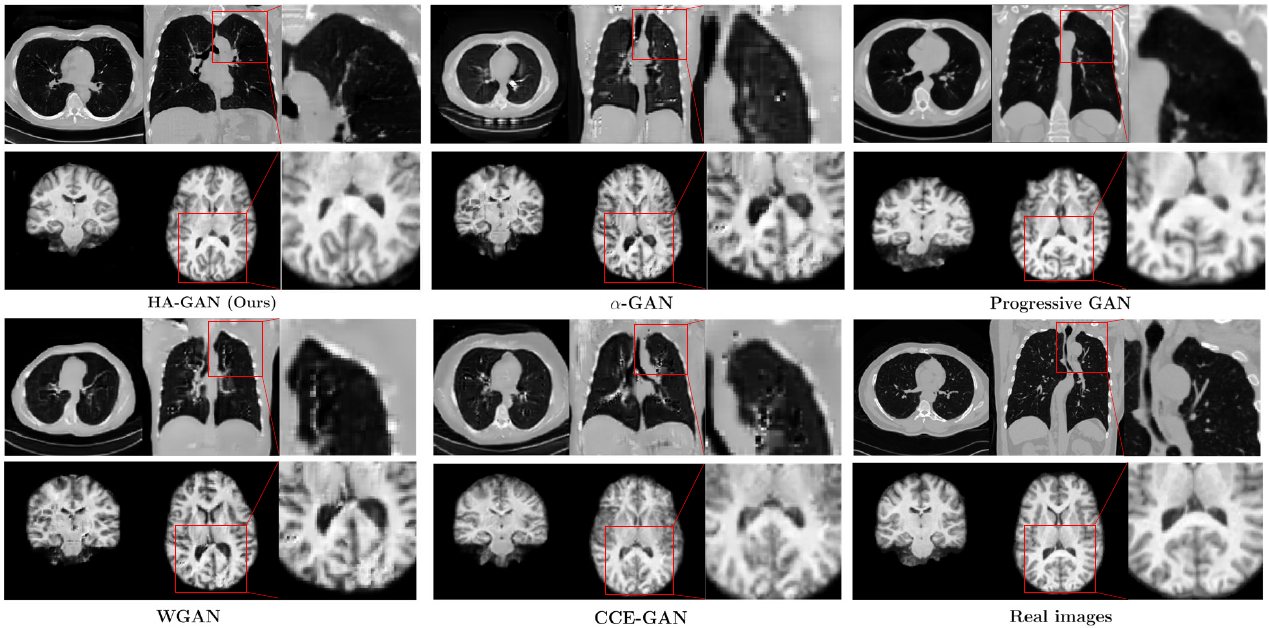}
    \caption[Caption]{ \emph{Randomly generated} images by different models and the real images. The figure illustrates that HA-GAN generates sharper images than the baselines.}
    \label{fig:result_synthesis}
\end{figure*}

To qualitatively analyze the results, we show some samples of synthetic images in Fig.~\ref{fig:result_synthesis}. The figure illustrates that HA-GAN generates sharper images than the baselines. 

\begin{figure}[t]
    \centering
    \begin{subfigure}{.35\textwidth}
    \includegraphics[width = 1.\textwidth]
    {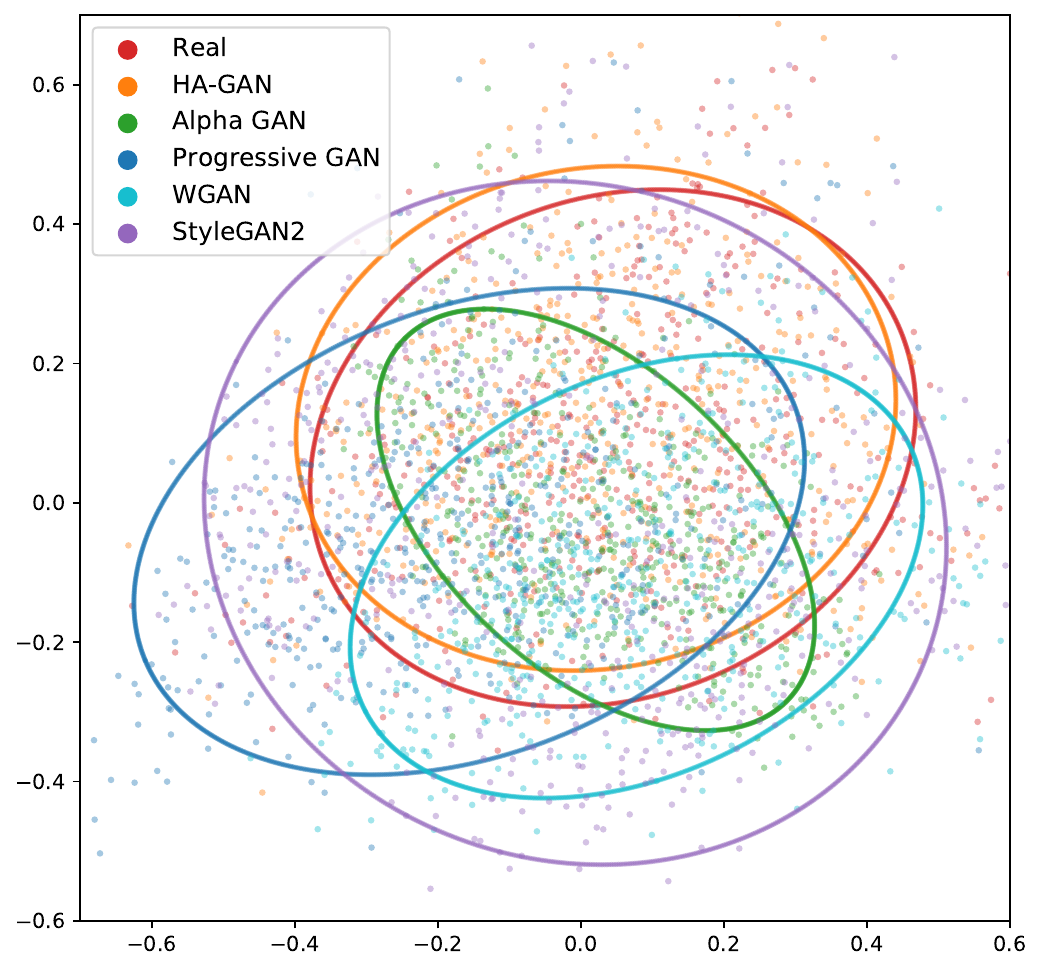}
    \caption[Caption]{ MDS visualization on COPDGene dataset.}
    \label{fig:result_pca_copd}
    \end{subfigure}
    \\
    \begin{subfigure}{.35\textwidth}
    \includegraphics[width = 1.\textwidth]
    {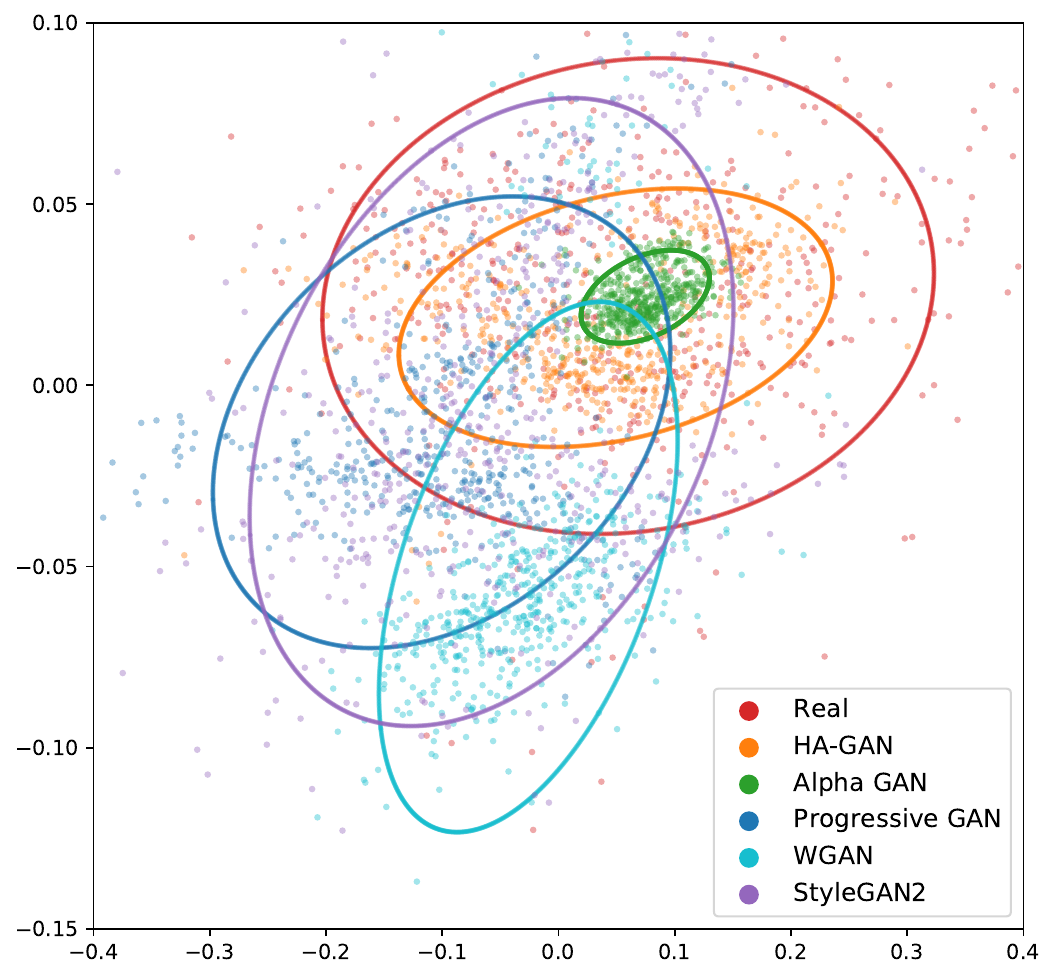}
    \caption[Caption]{ MDS visualization on GSP dataset. }
    \label{fig:result_pca_gsp}
    \end{subfigure}
    \caption{Comparison of the embedding of different models. We embed the features extracted from synthesized images into 2-dimensional space with MDS. The ellipses are fitted to scatters of each model for better visualization. The figures show that the embedding region of HA-GAN has the most overlapping with real images, compared to the baselines.}
\end{figure}

To examine the diversity and authenticity of generated images, we embed the synthetic and real images into the latent space. If the synthetic images are indistinguishable from the real images, then we expect that the synthetic and real images occupy the same region in the embedding space. Following the practice of~\cite{kwon2019generation}, we first use a pretrained 3D medical ResNet model~\cite{chen2019med3d} to extract features for 512 synthetic images by each method. As a reference, we also extract features for the real image samples using the same ResNet model. Then we conduct MDS to embed the exacted features into 2-dimensional space for both COPDGene and GSP datasets. The results are visualized in Fig. \ref{fig:result_pca_copd} and \ref{fig:result_pca_gsp}, respectively. To avoid cluttering dots, we only visualize four representative baseline methods. In both figures, we fit an ellipse for the embedding of each model with the least square. In the figures, we observe that synthetic images by HA-GAN better overlap with real images, compared with the baselines. This implies that HA-GAN generates more realistic-looking images than the baselines. 

\subsection{Data Augmentation for Supervised Learning}
In this experiment, we used the synthesized samples from HA-GAN to augment the training dataset for a supervised learning task.
Previous work~\cite{frid2018gan} has shown that GAN-generated samples improve the diversity of the training dataset, resulting in a better discriminative performance of the classifier.
Motivated by their results, we designed our experiment with the following three steps: First, we extended our HA-GAN architecture to enable conditional image generation and trained a class-conditional variant of HA-GAN. Next, we used trained HA-GAN to generate new images with class labels. Finally, we combined the original training dataset and GAN-generated images to train a multi-class classifier, and evaluate the performance on the test set.
We demonstrate our experiment on the COPDGene dataset using the GOLD score as a multi-class label. The GOLD score is a 5-class categorical variable ranging from 0-4.

We made two modifications to the original HA-GAN architecture to enable class-conditional image generation: 1) We updated the generator module $G^A(X;c)$ to take a one-hot code $c\sim p_c$ as input, along with latent variable $Z \sim \mathcal{N}( \mathbf{0}, \mathbf{I})$. $c$ represents the target class for the conditional image generation.
2) We updated the discriminator to output two probability distributions, one over the binary real/fake classification (same
as original HA-GAN), and another over the multi-class classification of class labels $P(C|X)$. Thus, the discriminator also acts as an auxiliary classifier for the class labels ~\cite{odena2017conditional}.
A schematic of the modified model can be found in Supplementary Material. In addition, two new terms are added to the original HA-GAN loss function for conditional generation:
\vspace{-1mm}
\begin{equation}
\label{Eq:loss_cgan}
\footnotesize
\begin{aligned}
    \mathcal{L}^H_{class}(G^H,G^A,D^H) = \E[\log P(C=c|X_r^H)] + \E[\log P(C=c|\widehat{X}_r^H)]
    \\
    \mathcal{L}^L_{class}(G^L,G^A,D^L) = \E[\log P(C=c|X^L)] + \E[\log P(C=c|\widehat{X}^L)]
\end{aligned}
\end{equation}

For comparison, we trained a class-conditional variant of $\alpha$-GAN on COPDGene dataset. The same two modifications discussed above are incorporated into the original $\alpha$-GAN model for conditional generation. We use a 3D CNN (implementation details are included in Supplementary Material Table.VIII) as the classification model.
We randomly sampled 80\% of subjects as training set and the rest are used as test set. We use an image size of $128^3$ for this experiment.
We divided $80\%$ of the subjects into training set, while the rest are included in a test set.
For creating the augmented training set, we combine randomly generated images from class-conditioned GAN (20\%) with the real images in the training set (80\%).
The proportion of different GOLD classes for generated images is the same as the original dataset.
We train two classifiers on the original training set and the GAN-augmented training set for 20 epochs respectively, and evaluated their performance on a held-out test set of real images.

Table~\ref{tbl:results_aug} shows the results on COPDGene dataset. Classifier trained with GAN augmented data performed better than the baseline model which trains on training set only consisted of  real images. Augmentation with HA-GAN can further improve performance compared to $\alpha$-GAN.
\begin{table}[htp]
\caption{Evaluation result for GAN-based data augmentation}
\centering
\begin{adjustbox}{max width=\textwidth}
\centering
\label{tbl:results_aug}
 \begin{tabular}{lc}
  \toprule
   Method  &  Accuracy(\%) \\
  \midrule
Baseline&$59.7$\\
Augmented with $\alpha$-GAN & $61.7$ \\
Augmented with HA-GAN & \bm{$62.9$}  \\
 \bottomrule
 \end{tabular}
\vspace{-3mm}
\end{adjustbox}
\end{table}

\subsection{Clinical-Relevant Feature Extraction}
In this section, we evaluate the encoded latent variables from real images to predict clinical-relevant measurements. This task evaluates how much information about the disease severity is preserved in the encoded latent features.

\begin{table}[t]
\centering
\caption{$R^2$ for predicting clinical-relevant measurements}
\label{tbl:r2}
\begin{tabular}{lccc}

\toprule
Method
& $\log$ \texttt{FEV1pp} 
& $\log$ \texttt{$\text{FEV}_1 / \text{FVC}$} 
& $\log$ \texttt{$\%$Emphysema} 
\\
\midrule
VAE-GAN
& 0.215 & 0.315 & 0.375
\\
$\alpha$-GAN
& 0.512 & 0.622 & 0.738
\\
\midrule
HA-GAN
& \bf 0.555 & \bf 0.657 & \bf 0.746 
\\
\bottomrule
\multicolumn{4}{p{.45\textwidth}}{We do not include the results of WGAN and Progressive GAN, because they do not incorporate an encoder.}
 \end{tabular}
\vspace{-4mm}
\end{table}

We select two respiratory measurements and one CT-based measurement of emphysema to measure disease severity. For respiratory measurements, we use percent predicted values of Forced Expiratory Volume in one second (\texttt{FEV1pp}) and its ratio with Forced vital capacity (FVC) (\texttt{$\text{FEV}_1 / \text{FVC}$}).   
Given extracted features, we train a Ridge regression model with $\lambda = 1\times10^{-4}$ to predict the {\it logarithm} of each of the measurements. We report the $R^2$ scores on held-out test data. Table~\ref{tbl:r2} shows that HA-GAN achieves higher $R^2$ than the baselines. 
The results imply that HA-GAN preserves more information about the disease severity than baselines.

\subsection{Exploring the Latent Space}
\begin{figure}[t]
    \centering
    \includegraphics[width=.48\textwidth ]{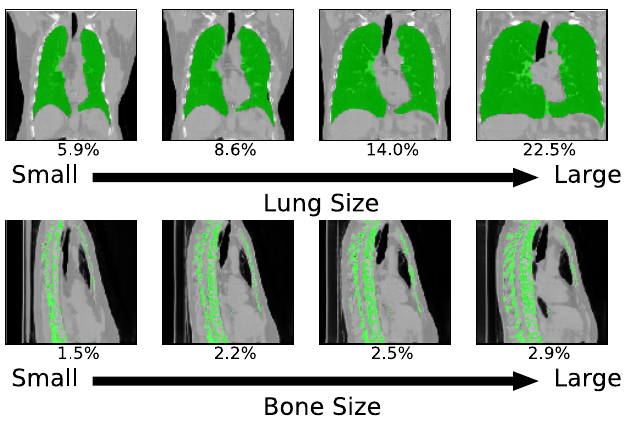}
    \caption{ Latent space exploration on thorax CT images. 
    The figure reports synthetic images generated by changing the latent code in two different directions, corresponding to the lung and bone volume respectively. The number shown below each slice indicates the percentage of the volume of interest that occupies the volume of lung region of the synthetic image. The segmentation masks are plotted in green.}
    \label{fig:exploring}
\end{figure}

This section investigates whether change along a certain direction in the latent space corresponds to semantic meanings. 
We segment the lung regions in the thorax CT images using Chest Image Platform (CIP)\cite{san2015chest}, and segment the bone tissues via thresholding. The detailed thresholding criteria can be found in Supplementary Material.
Next, we train linear regression models that predict the total volume of the different tissues/regions with the encoded latent representations $ Z $ for each image, optimizing with least square. The learned parameter vector for each class represents the latent direction.
Then, we manipulate the latent variable along the direction corresponding to the learned parameters of linear models and generate the images by feeding the resulted latent representations into the generator. More specifically, first a reference latent variable is randomly sampled, then the latent variable is moved along the latent direction learned until the target volume is reached, which is predicted by the linear regression model. As shown in Fig. \ref{fig:exploring}, for thorax CT images, we identify directions in latent space corresponding to the volume of lung and bone respectively. 
When we go along these directions in latent space, we can observe the change of volumes for these tissues.

\subsection{Memory Efficiency}
\begin{figure}[t]
    \centering
    \includegraphics[trim=0 0 0 20, clip, width = 0.5\textwidth]
    {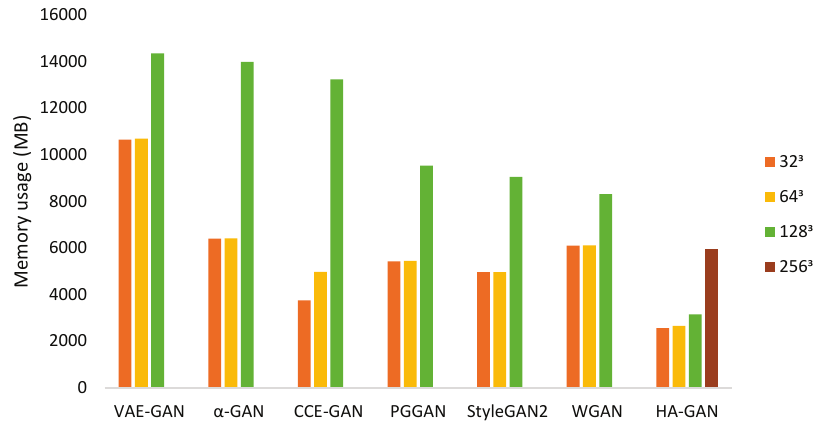}
    \caption[Caption]{ Results of memory usage test. 
    Note that HA-GAN is the only model that can generate images sized $256^3$ without memory overflow on high-end GPU with 16GB VRAM.
    }
    \label{fig:result_memory}
\end{figure}

In this section, we compare the memory efficiency of HA-GAN with baselines. We measure the GPU memory usage at the training time for all models under different resolutions, including $32^3$, $64^3$, $128^3$, and $256^3$. The results are shown in Fig.~\ref{fig:result_memory}. Note that the experiments are performed on the same GPU (Tesla V100 with 16GB memory), and we set the batch size to 2. The HA-GAN consumes much less memory than baseline models under different resolutions. In addition, HA-GAN is the only model that can generate images of sizes $256^3$. All other models exhaust the entire memory of GPU; thus, the memory demand cannot be measured.
In order to investigate where the memory efficiency comes from, we report the number of parameters for HA-GAN at different resolutions in Table~\ref{tbl:parameter}. We found that as the resolution increases, the number of parameters only increases marginally, which is expected as the model only requires a few more layers as resolution increases. 

In addition, we compare the computational efficiency of our HA-GAN model with baseline models. More specifically, we measure the number of iterations per second during training. One NVIDIA Tesla V100 GPU is used for each model and we set the batch size as 2. The comparison is performed under the $128^3$ resolution where all models can fit in memory. The result is shown in Table~\ref{tbl:computation}. Our HA-GAN is more computationally efficient than the baselines.

\begin{table}[htp]
\caption{Number of model parameters and memory usage under different resolutions}
\centering
\begin{adjustbox}{max width=\textwidth}
\centering
\label{tbl:parameter}
\begin{tabular}{lcc}
   \toprule
  Output Resolution&Memory Usage (MB)&\#Parameters\\
  \midrule
$32^3$&2573&74.7M\\
$64^3$&2665&78.7M\\
$128^3$&3167&79.6M\\
$256^3$&5961&79.7M\\
 \bottomrule
 \end{tabular}
\vspace{-3mm}
\end{adjustbox}
\end{table}

\begin{table}[htp]
\caption{Training speed (iter/s) for different models (Higher is better)}
\centering
\begin{adjustbox}{max width=.5\textwidth}
\label{tbl:computation}
\begin{tabular}{ccccccc}
\toprule
WGAN&VAE-GAN&PGGAN&$\alpha$-GAN&CCE-GAN&StyleGAN&HA-GAN
\\
\midrule
2.0 & 1.0 & 1.3 & 1.6 & 0.35& 0.23&\textbf{3.8}
\\
\bottomrule
\end{tabular}
\vspace{-3mm}
\end{adjustbox}
\end{table}

\section{Discussion}
As shown quantitatively in Table~\ref{tbl:FID_COPD} and Table~\ref{tbl:FID_GSP}, HA-GAN achieves lower FID and MMD, as well as higher IS. This implies that our model generates more realistic images. This is further confirmed by the synthetic images shown in Fig. \ref{fig:result_synthesis}, where HA-GAN generates sharper images compared to other methods. We found that our method outperforms baseline methods at both the resolution of $128^3$ and $256^3$, but the lead is larger at $256^3$ resolution than $128^3$. 
Based on the results, we believe that the sharp generation results come from both the model itself and its ability to directly generate images at $256^3$ without interpolation upsampling.
For the baseline models, we found that $\alpha$-GAN and WGAN have similar performance, and VAE-GAN tends to generate blurry images. WGAN is essentially the $\alpha$-GAN without the encoder. Based on qualitative examples shown in Fig. 3, it can generate sharper images compared to $\alpha$-GAN and Progressive GAN. However, it also generates more artifacts.
According to the quantitative analysis shown in Table 2, overall the generation quality of $\alpha$-GAN is comparable with WGAN. Although our proposed HA-GAN achieves the highest quality comparing to the baseline models, we admit that there is still a gap between HA-GAN generated images and real images.
We also note that in order to achieve optimal performance for HA-GAN, most of blank axial slices of training images need to be removed, because blank sub-volume may confuse the model.
There are several directions that may further improve the performance, including using a pretrained segmentation network to regularize the generated images, adding an attention layer to the generator, etc. We hope that our method establishes a strong baseline that can be pushed further by future work.

For the ablation studies, first we found that adding a low-resolution branch can help improve results, we think it's because the low-resolution branch can help the model learn the global structure.
Second, we observe in Table \ref{tbl:ablation} that HA-GAN with encoder outperforms the version without encoder in terms of image synthesis quality. 
The reconstruction loss in the objective function ensures that the reconstructed images are voxel-wise consistent with the original images. This term can encourage the generator to represent all data and not collapse, improving the performance of the generator in terms of image synthesis. 
Finally, using randomly selected $r$ leads to randomly selected locations of sub-volumes. In this way, the junctions between sub-volumes can be better covered.
 
The embedding shown in Fig.~\ref{fig:result_pca_copd} and Fig.~\ref{fig:result_pca_gsp} reveals that the distribution of the synthetic images by HA-GAN is more consistent with the real images, compared to all baselines. 
The scatters of WGAN/$\alpha$-GAN show compressed support of real data distribution, which suggests that samples of WGAN (cyan) and $\alpha$-GAN (green) have lower diversity than the real images.
We think one reason is that the models only learn few attributes of samples in the dataset. 
To be more specific, the models learn an overly simplified distribution, so the generated images are of lower diversity.
The HA-GAN model we proposed has an encoder module, which encourages different latent codes to map to different outputs, improving the diversity of generated samples.
A portion of scatters of Progressive GAN (blue) and StyleGAN2 (purple) lay outside of real data distribution (red), which suggests that some generated images may contain artifacts.

In clinical applications, high-resolution CT can help radiologists make reliable diagnose decisions, including pulmonary eosinophilic granuloma, lymphangiomyomatosis, and emphysema~\cite{bonelli1998accuracy}. 
High-resolution CT is especially beneficial in imaging tasks in which small anatomy and pathologic structure is the target, such as in-stent stenosis, lung nodules, coronary calcification, and temporal bones~\cite{doi:10.1148/radiol.2018181156}.
There are previous works that propose to use 3D GAN for diverse clinical applications~\cite{cirillo2020vox2vox,yu20183d}. For instance, synthesized images can be used for data anonymization which enables privacy-preserving data sharing between institutions~\cite{subramaniam2022generating}. However, the generated images are limited to the small size of $128 \times 128 \times 128$ or below, due to insufficient memory during training.
In most clinical CT applications, image matrix size of $512 \times 512$ or larger is used for in-plane direction~\cite{doi:10.1148/radiol.2018181156}.
Our proposed HA-GAN bridges the gap between them and serve as a plug-and-play module to improve performance for many GAN-based medical imaging applications.

We demonstrate two clinical applications in our paper: data augmentation and clinical-relevant feature extraction.
For data augmentation, the results in Table.~\ref{tbl:results_aug} show that samples generated by HA-GAN can help the training of classification model. While samples generated by $\alpha$-GAN can also help the training, the performance gain is smaller. We think one reason is that samples generated by HA-GAN are more realistic, also shown in Table~\ref{tbl:FID_COPD} and Table~\ref{tbl:FID_GSP}. GAN can learn a rich prior from existing medical imaging datasets, and the generated samples can help classifiers to achieve better performance.

For the experiment of feature extraction, we encode the full image into a flat variable to extract meaningful and compact feature representation for downstream clinical feature prediction.
Table \ref{tbl:r2} shows that HA-GAN can better extract clinical-relevant features from the images, comparing to VAE-GAN and $\alpha$-GAN. Some clinical-relevant information might be hidden in specific details in the medical images, and can only be observed under high resolution. VAE-GAN and $\alpha$-GAN can only process lower-resolution images of $128^3$. We speculate that the high-resolution information leveraged by HA-GAN helps it learn better representations.

From Table.~\ref{tbl:parameter}, we found that as the output resolution increases, the total number of model parameters does not increase much,
 but as the multiplier factor increases, the memory usage increases drastically. 
 Therefore, we believe that the memory efficiency mainly comes from the sub-volume scheme rather than model parameters.

\section{Conclusion}
In this work, we develop a hierarchical GAN model that can generate 3D high-resolution images. 
Experiments on 3D thorax CT and brain MRI show that HA-GAN achieves state-of-the-art performance in image synthesis and clinical applications. 
Our method enables various real-world medical imaging applications that rely on high-resolution image generation and analysis.

\bibliography{egbib.bib}
\bibliographystyle{IEEEtran}

\end{document}


\markboth{\journalname, VOL. XX, NO. XX, XXXX 2021}
{L. Sun \MakeLowercase{\textit{et al.}}: Preparation of Papers for IEEE TRANSACTIONS and JOURNALS (Nov 2021)}

\title{Supplementary Material\vspace{-1.4em}}
%
%
\author{}
%
\maketitle              
\hspace{-3.5mm}\textbf{Network architecture}$\quad$
In the tables below, we show the detailed architecture of network modules of HA-GAN, including $G^A$, $G^L$, $G^H$, $D^L$, $D^H$, $E^H$ and $E^G$.
\begin{table}[H]
\centering
\caption{Architecture of the $G^A$ Network}
\label{tbl:arch_ga}
\begin{adjustbox}{max width=0.38\textwidth}
\begin{tabular}{lcc}

\toprule
Layer
&Filter size, stride
&Output size$(C,D,H,W)$
\\
\midrule
Input
& - & 1 $\times$ 1024
\\
Dense
& - & 512$\times$4$\times$4$\times$4
\\
\midrule
Conv3D
& 3$\times$3$\times$3, 1 & 512$\times$4$\times$4$\times$4
\\
GroupNorm+ReLU
& - & 512$\times$4$\times$4$\times$4
\\
Interpolation
& - & 512$\times$8$\times$8$\times$8
\\
\midrule
Conv3D
& 3$\times$3$\times$3, 1 & 512$\times$8$\times$8$\times$8
\\
GroupNorm+ReLU
& - & 512$\times$8$\times$8$\times$8
\\
Interpolation
& - & 512$\times$16$\times$16$\times$16
\\
\midrule
Conv3D
& 3$\times$3$\times$3, 1 & 256$\times$16$\times$16$\times$16
\\
GroupNorm+ReLU
& - & 256$\times$16$\times$16$\times$16
\\
Interpolation
& - & 256$\times$32$\times$32$\times$32
\\
\midrule
Conv3D
& 3$\times$3$\times$3, 1 & 128$\times$32$\times$32$\times$32
\\
GroupNorm+ReLU
& - & 128$\times$32$\times$32$\times$32
\\
Interpolation
& - & 128$\times$64$\times$64$\times$64
\\
\midrule
Conv3D
& 3$\times$3$\times$3, 1 & 64$\times$64$\times$64$\times$64
\\
GroupNorm+ReLU
& - & 64$\times$64$\times$64$\times$64
\\
\bottomrule
\end{tabular}
\end{adjustbox}
\end{table}

\begin{table}[H]
\centering
\caption{Architecture of the $G^L$ Network}
\label{tbl:arch_gl}
\begin{adjustbox}{max width=0.38\textwidth}
\begin{tabular}{lcc}

\toprule
Layer
&Filter size, stride
&Output size$(C,D,H,W)$
\\
\midrule
Input
& - & 64$\times$64$\times$64$\times$64
\\
\midrule
Conv3D
& 3$\times$3$\times$3, 1 & 32$\times$64$\times$64$\times$64
\\
GroupNorm+ReLU
& - & 32$\times$64$\times$64$\times$64
\\
\midrule
Conv3D
& 3$\times$3$\times$3, 1 & 16$\times$64$\times$64$\times$64
\\
GroupNorm+ReLU
& - & 16$\times$64$\times$64$\times$64
\\
\midrule
Conv3D
& 3$\times$3$\times$3, 1 & 1$\times$64$\times$64$\times$64
\\
Tanh
& - & 1$\times$64$\times$64$\times$64
\\
\bottomrule
\end{tabular}
\end{adjustbox}
\end{table}

\begin{table}[H]
\centering
\caption{Architecture of the $G^H$ Network}
\label{tbl:arch_gh}
\begin{adjustbox}{max width=0.38\textwidth}
\begin{tabular}{lcc}

\toprule
Layer
&Filter size, stride
&Output size$(C,D,H,W)$
\\
\midrule
Input
& - & 64$\times$64$\times$64$\times$64
\\
Interpolation
& - & 64$\times$128$\times$128$\times$128
\\
\midrule
Conv3D
& 3$\times$3$\times$3, 1 & 32$\times$128$\times$128$\times$128
\\
GroupNorm+ReLU
& - & 32$\times$128$\times$128$\times$128
\\
Interpolation
& - & 32$\times$256$\times$256$\times$256
\\
\midrule
Conv3D
& 3$\times$3$\times$3, 1 & 1$\times$256$\times$256$\times$256
\\
Tanh
& - & 1$\times$256$\times$256$\times$256
\\
\bottomrule
\end{tabular}
\end{adjustbox}
\end{table}

\begin{table}[H]
\centering
\caption{Architecture of the $E^H$ Network}
\label{tbl:arch_eh}
\begin{adjustbox}{max width=0.38\textwidth}
\begin{tabular}{lcc}

\toprule
Layer
&Filter size, stride
&Output size$(C,D,H,W)$
\\
\midrule
Input
& - & 1$\times$32$\times$256$\times$256
\\
\midrule
Conv3D
& 4$\times$4$\times$4, 2 & 32$\times$16$\times$128$\times$128
\\
GroupNorm+ReLU
& - & 32$\times$16$\times$128$\times$128
\\
\midrule
Conv3D
& 3$\times$3$\times$3, 1 & 32$\times$16$\times$128$\times$128
\\
GroupNorm+ReLU
& - & 32$\times$16$\times$128$\times$128
\\
\midrule
Conv3D
& 4$\times$4$\times$4, 2 & 64$\times$8$\times$64$\times$64
\\
GroupNorm+ReLU
& - & 64$\times$8$\times$64$\times$64
\\
\bottomrule
\end{tabular}
\end{adjustbox}
\end{table}

\begin{figure}[htp]
\centering   
\includegraphics[width=0.36\textwidth]{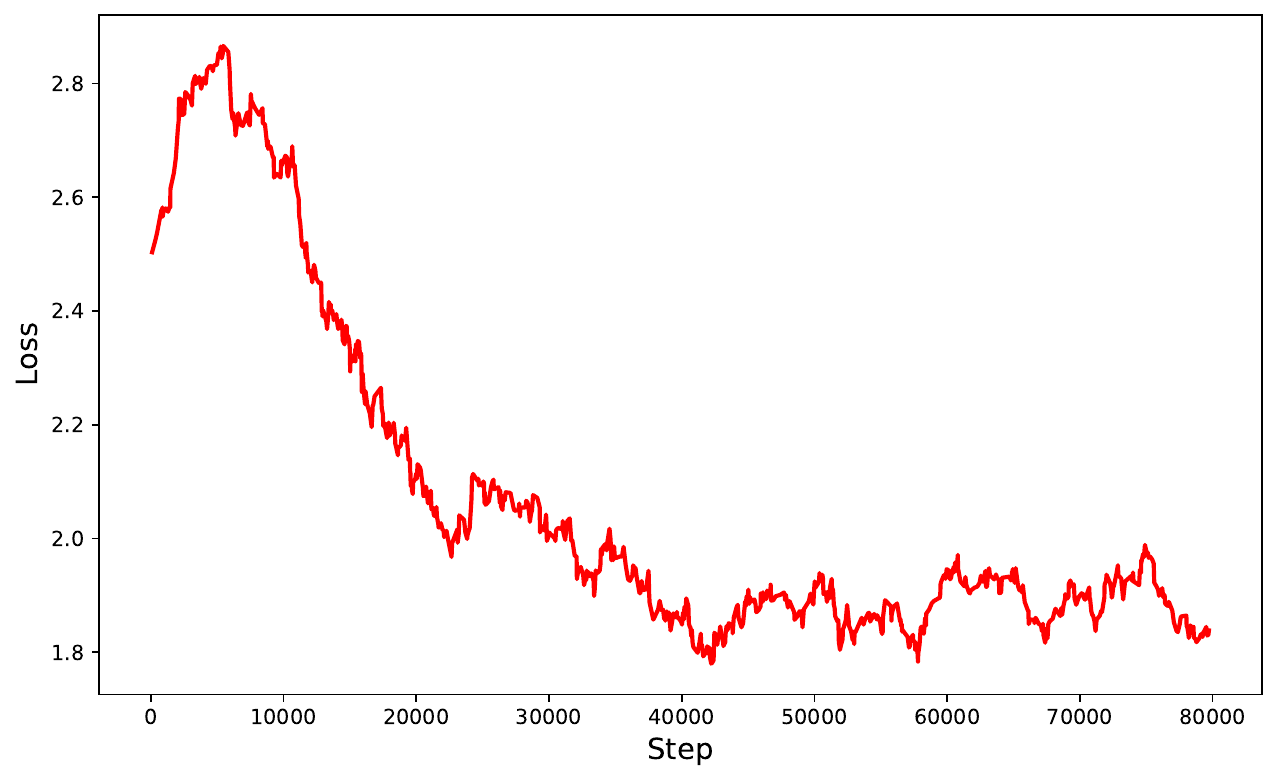}
\caption{Training loss curve of the generator for HA-GAN.}
\label{fig:curve_g}
\vspace{2mm}
\centering   
\includegraphics[width=0.36\textwidth]{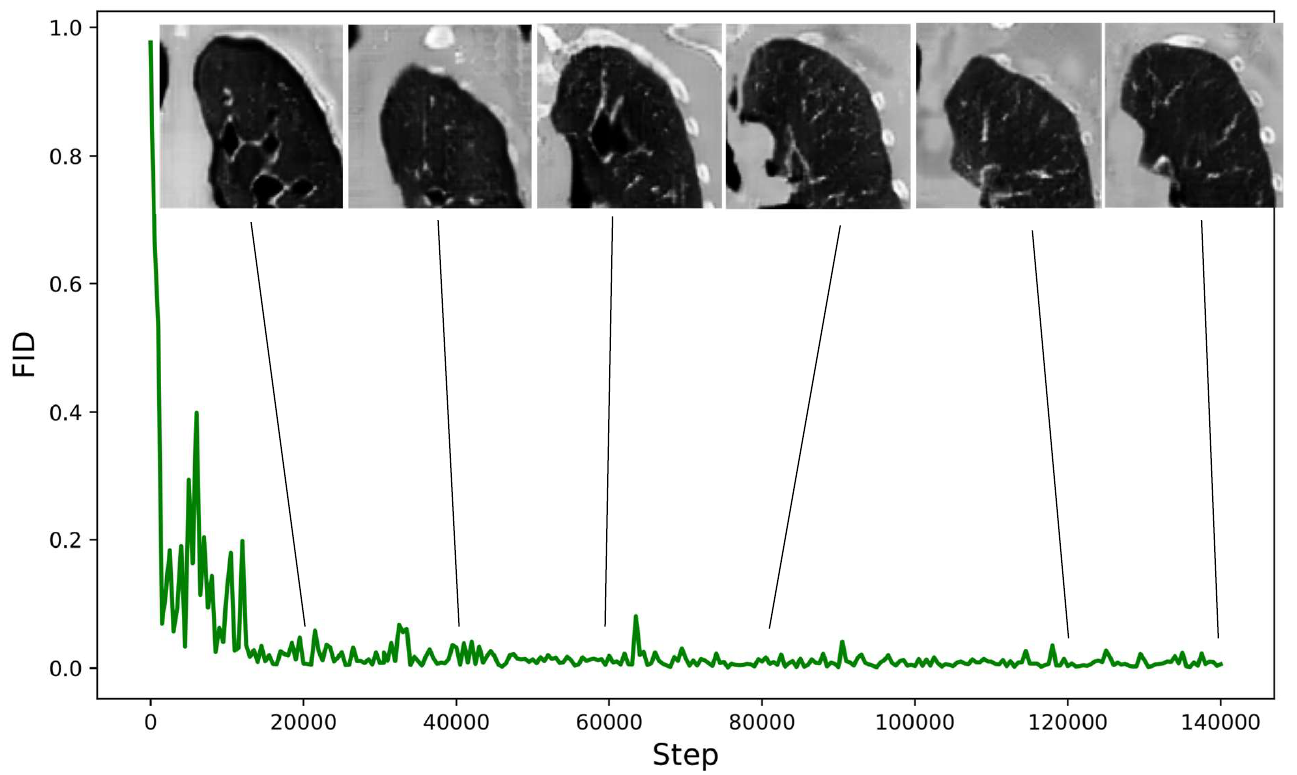}
\caption{Validation curve during training for HA-GAN. FID score is used as the validation metric. The images cropped on left upper lobe shown above demonstrate the quality of random image synthesis during training.}
\label{fig:curve_fid}

\centering   
\includegraphics[width=0.39\textwidth]{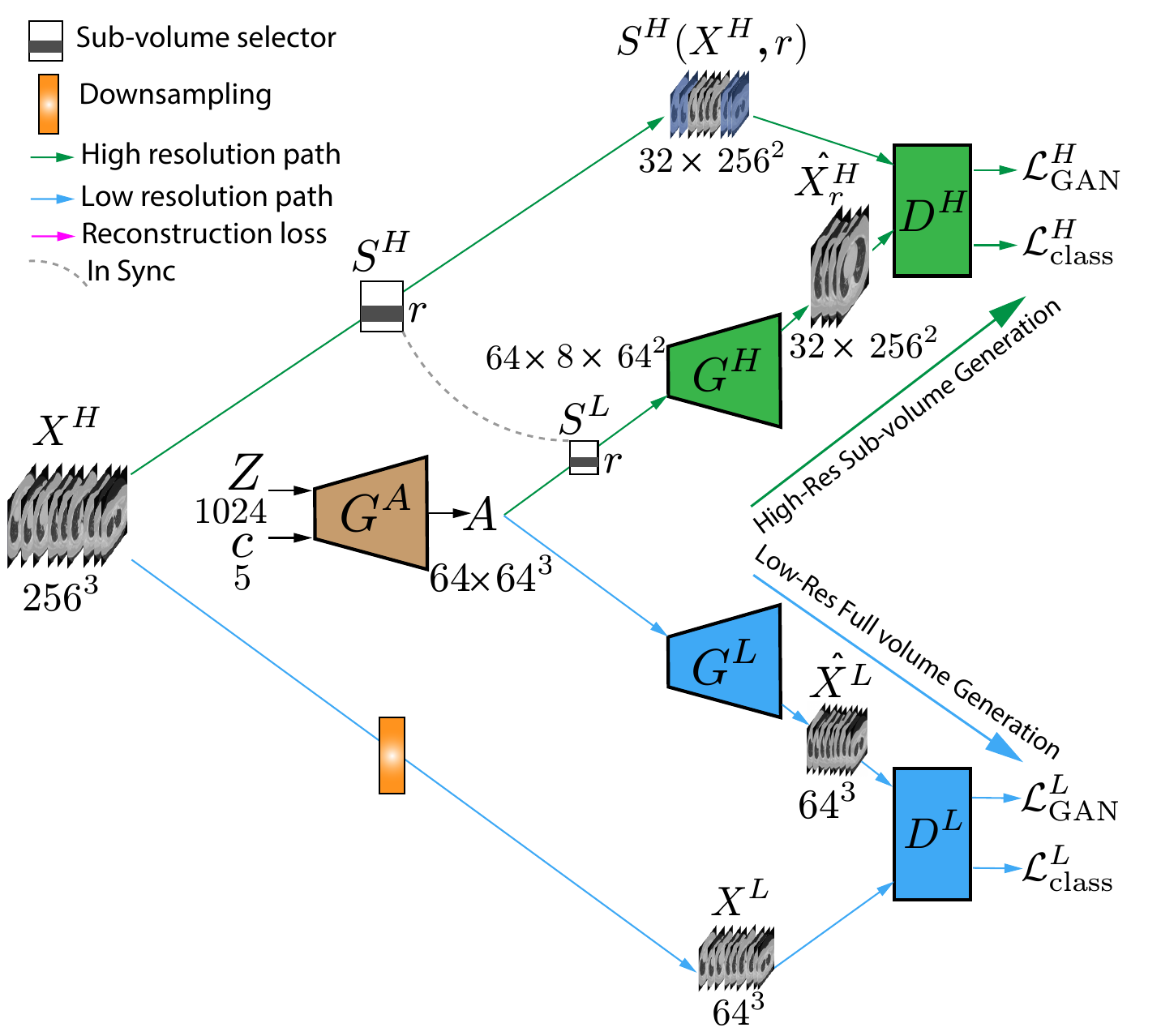}
\caption{Architecture of conditional HA-GAN for data augmentation (encoder is hidden here to improve clarity).}
\label{fig:arch_acgan}
\end{figure}

\begin{table}[H]
\centering
\caption{Architecture of the $E^G$ Network}
\label{tbl:arch_el}
\begin{adjustbox}{max width=0.38\textwidth}
\begin{tabular}{lcc}

\toprule
Layer
&Filter size, stride
&Output size$(C,D,H,W)$
\\
\midrule
Input
& - & 64$\times$64$\times$64$\times$64
\\
\midrule
Conv3D
& 4$\times$4$\times$4, 2 & 32$\times$32$\times$32$\times$32
\\
GroupNorm+ReLU
& - & 32$\times$32$\times$32$\times$32
\\
\midrule
Conv3D
& 4$\times$4$\times$4, 2 & 64$\times$16$\times$16$\times$16
\\
GroupNorm+ReLU
& - & 64$\times$16$\times$16$\times$16
\\
\midrule
Conv3D
& 4$\times$4$\times$4, 2 & 128$\times$8$\times$8$\times$8
\\
GroupNorm+ReLU
& - & 128$\times$8$\times$8$\times$8
\\
\midrule
Conv3D
& 4$\times$4$\times$4, 2 & 256$\times$4$\times$4$\times$4
\\
GroupNorm+ReLU
& - & 256$\times$4$\times$4$\times$4
\\
\midrule
Conv3D
& 4$\times$4$\times$4, 1 & 1024$\times$1$\times$1$\times$1
\\
\bottomrule
\end{tabular}
\end{adjustbox}
\end{table}
\vspace{-6mm}
\begin{table}[H]
\centering
\caption{Architecture of the $D^L$ Network}
\label{tbl:arch_dl}
\begin{adjustbox}{max width=0.38\textwidth}
\begin{tabular}{lcc}

\toprule
Layer
&Filter size, stride
&Output size$(C,D,H,W)$
\\
\midrule
Input
& - & 1$\times$64$\times$64$\times$64
\\
\midrule
Conv3D
& 4$\times$4$\times$4, 2 & 32$\times$32$\times$32$\times$32
\\
SpectralNorm+LeakyReLU
& - & 32$\times$32$\times$32$\times$32
\\
\midrule
Conv3D
& 4$\times$4$\times$4, 2 & 64$\times$16$\times$16$\times$16
\\
SpectralNorm+LeakyReLU
& - & 64$\times$16$\times$16$\times$16
\\
\midrule
Conv3D
& 4$\times$4$\times$4, 2 & 128$\times$8$\times$8$\times$8
\\
SpectralNorm+LeakyReLU
& - & 128$\times$8$\times$8$\times$8
\\
\midrule
Conv3D
& 4$\times$4$\times$4, 2 & 256$\times$4$\times$4$\times$4
\\
SpectralNorm+LeakyReLU
& - & 256$\times$4$\times$4$\times$4
\\
\midrule
Conv3D
& 4$\times$4$\times$4, 1 & 1$\times$1$\times$1$\times$1
\\
Reshape
& - & 1
\\
\bottomrule
\end{tabular}
\end{adjustbox}
\end{table}
\vspace{-4mm}
\begin{table}[H]
\centering
\caption{Architecture of the $D^H$ Network}
\label{tbl:arch_dh}
\begin{adjustbox}{max width=0.38\textwidth}
\begin{tabular}{lcc}

\toprule
Layer
&Filter size, stride
&Output size$(C,D,H,W)$
\\
\midrule
Input
& - & 1$\times$32$\times$256$\times$256
\\
\midrule
Conv3D
& 4$\times$4$\times$4, 2 & 16$\times$16$\times$128$\times$128
\\
SpectralNorm+LeakyReLU
& - & 16$\times$16$\times$128$\times$128
\\
\midrule
Conv3D
& 4$\times$4$\times$4, 2 & 32$\times$8$\times$64$\times$64
\\
SpectralNorm+LeakyReLU
& - & 32$\times$8$\times$64$\times$64
\\
\midrule
Conv3D
& 4$\times$4$\times$4, 2 & 64$\times$4$\times$32$\times$32
\\
SpectralNorm+LeakyReLU
& - & 64$\times$4$\times$32$\times$32
\\
\midrule
Conv3D
& 2$\times$4$\times$4, 2 & 128$\times$2$\times$16$\times$16
\\
SpectralNorm+LeakyReLU
& - & 128$\times$2$\times$16$\times$16
\\
\midrule
Conv3D
& 2$\times$4$\times$4, 1 & 256$\times$1$\times$8$\times$8
\\
SpectralNorm+LeakyReLU
& - & 256$\times$1$\times$8$\times$8
\\
\midrule
Conv3D
& 1$\times$4$\times$4, 1 & 512$\times$1$\times$4$\times$4
\\
SpectralNorm+LeakyReLU
& - & 512$\times$1$\times$4$\times$4
\\
\midrule
Conv3D
& 1$\times$4$\times$4, 1 & 128$\times$1$\times$1$\times$1
\\
SpectralNorm+LeakyReLU
& - & 128$\times$1$\times$1$\times$1
\\
\midrule
Dense
& - & 64
\\
SpectralNorm+LeakyReLU
& - & 64
\\
\midrule
Dense
& - & 32
\\
SpectralNorm+LeakyReLU
& - & 32
\\
\midrule
Dense
& - & 1
\\
\bottomrule
\end{tabular}
\end{adjustbox}
\end{table}
\vspace{-2mm}
\hspace{-3.5mm}\textbf{Datasets and Training Details}$\quad$
For the COPDGene dataset, CT scans are acquired using multi-detector CT scanners (at least 16 detector channels).
Volumetric CT acquisitions are obtained on full inspiration (200mAs). The initial voxel size is typically $0.68mm\times0.68mm\times0.54mm$. The dataset is available at \url{https://www.ncbi.nlm.nih.gov/projects/gap/cgi-bin/study.cgi?study_id=phs000179.v6.p2}
For the GSP dataset, all imaging data were collected on matched 3T Tim Trio scanners (Siemens Healthcare, Erlangen, Germany) at Harvard University and Massachusetts General Hospital using the vendor-supplied 12-channel phased-array head coil. Structural data included a high-resolution (1.2mm isotropic) multi-echo T1-weighted magnetization-prepared gradient-echo image. 
The dataset is available at \url{https://dataverse.harvard.edu/dataverse/GSP}
For fair comparison between different baseline models, we use the official implementation whenever it is available.
We also provide training and validation curves in Fig.~\ref{fig:curve_g} and Fig.~\ref{fig:curve_fid}.

\vspace{-1mm}
\hspace{-2.5mm}\textbf{Data Augmentation for Supervised Learning}$\quad$
In this section, we provide more details about our experiment of data augmentation for supervised learning. In Fig.~\ref{fig:arch_acgan}, we show the architecture of conditional HA-GAN for data augmentation. Compared with original HA-GAN, two modifications are made to the original HA-GAN architecture: 1) Besides the latent variable $Z$ sampled from Gaussian distribution, the generator module $G^A(Z;c)$ also takes a 5-class one-hot code variable $c\sim p_c$ as input, which indicates which class the generated images should be. 2) The discriminator gives both a probability distribution over real/fake classification (same as original HA-GAN) and a probability distribution over the class labels $P(C|X)$. In this way, the discriminator also serves as auxiliary classifier for GOLD score. The network architecture of the 3D CNN used for supervised training is shown in~\ref{tbl:arch_3dgan}.

\begin{table}[H]
\centering
\caption{Architecture of the 3D CNN Network}
\label{tbl:arch_3dgan}
\begin{adjustbox}{max width=0.34\textwidth}
\begin{tabular}{lcc}

\toprule
Layer
&Filter size, stride
&Output size$(C,D,H,W)$
\\
\midrule
Input
& - & 1$\times$128$\times$128$\times$128
\\
\midrule
Conv3D
& 3$\times$3$\times$3, 1 & 8$\times$128$\times$128$\times$128
\\
BatchNorm+ELU
& - & 8$\times$128$\times$128$\times$128
\\
\midrule
Conv3D
& 3$\times$3$\times$3, 2 & 8$\times$64$\times$64$\times$64
\\
BatchNorm+ELU
& - & 8$\times$64$\times$64$\times$64
\\
\midrule
Conv3D
& 3$\times$3$\times$3, 1 & 16$\times$64$\times$64$\times$64
\\
BatchNorm+ELU
& - & 16$\times$64$\times$64$\times$64
\\
Conv3D
& 3$\times$3$\times$3, 1 & 16$\times$64$\times$64$\times$64
\\
BatchNorm+ELU
& - & 16$\times$64$\times$64$\times$64
\\
Conv3D
& 3$\times$3$\times$3, 2 & 16$\times$32$\times$32$\times$32
\\
BatchNorm+ELU
& - & 16$\times$32$\times$32$\times$32
\\
\midrule
Conv3D
& 3$\times$3$\times$3, 1 & 32$\times$32$\times$32$\times$32
\\
BatchNorm+ELU
& - & 32$\times$32$\times$32$\times$32
\\
Conv3D
& 3$\times$3$\times$3, 1 & 32$\times$32$\times$32$\times$32
\\
BatchNorm+ELU
& - & 32$\times$32$\times$32$\times$32
\\
Conv3D
& 3$\times$3$\times$3, 2 & 32$\times$16$\times$16$\times$16
\\
BatchNorm+ELU
& - & 32$\times$16$\times$16$\times$16
\\
\midrule
Conv3D
& 3$\times$3$\times$3, 1 & 64$\times$16$\times$16$\times$16
\\
BatchNorm+ELU
& - & 64$\times$16$\times$16$\times$16
\\
Conv3D
& 3$\times$3$\times$3, 1 & 64$\times$16$\times$16$\times$16
\\
BatchNorm+ELU
& - & 64$\times$16$\times$16$\times$16
\\
Conv3D
& 3$\times$3$\times$3, 2 & 64$\times$8$\times$8$\times$8
\\
BatchNorm+ELU
& - & 64$\times$8$\times$8$\times$8
\\
\midrule
Conv3D
& 3$\times$3$\times$3, 1 & 128$\times$8$\times$8$\times$8
\\
BatchNorm+ELU
& - & 128$\times$8$\times$8$\times$8
\\
\midrule
Conv3D
& 3$\times$3$\times$3, 2 & 128$\times$4$\times$4$\times$4
\\
BatchNorm+ELU
& - & 128$\times$4$\times$4$\times$4
\\
\midrule
AvgPool
& - & 128$\times$1$\times$1$\times$1
\\
Reshape
& - & 128
\\
\midrule
Dense
& - & 5
\\
\bottomrule
\end{tabular}
\end{adjustbox}

\caption{Testing the impact of sub-volume multiplier }
\centering
\begin{adjustbox}{max width=\textwidth}
\centering
\label{tbl:multiplier}
\begin{tabular}{lc}
   \toprule
  Multiplier Factor&Memory Usage (MB)\\
  \midrule
$1 / 8$&5961\\
$1 / 4$&10689\\
$1 /2 $&13185\\
 \bottomrule
 \end{tabular}
\vspace{-3mm}
\end{adjustbox}
\end{table}

\begin{figure}[htp]
\centering   
\includegraphics[width=.42\textwidth]{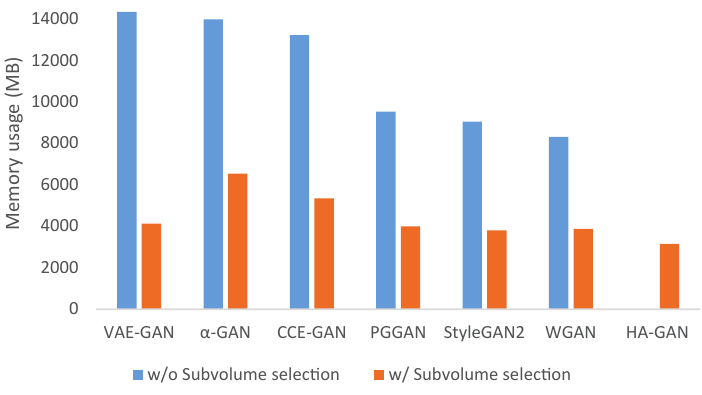}
\vspace{-0.5mm}
\caption{Effect of Sub-volume selection on memory usage}
\label{fig:memory_subvolume}
\end{figure}

\hspace{-5mm}\textbf{Effect of Sub-volume selection on memory usage}$\quad$ We split the baseline models by applying our proposed sub-volume selection method to the baseline models, and measured memory usage during training. The output size of the original generator for each model is set as $128^3$, and we set the size of sub-volume as $32\times128^2$. The batch size is set as 2. The results are shown in Fig.~\ref{fig:memory_subvolume}. We can see that our proposed sub-volume selection method drastically reduces the memory demand during training for baseline methods.
\vspace{-4mm}
